\documentclass[10pt,conference]{IEEEtran}
\usepackage{amsmath,amsfonts,graphicx,hyperref}

\usepackage[natbib, bibencoding=utf8, citestyle=numeric, bibstyle=ieee, maxbibnames=999, maxcitenames=2, mincitenames=1, sortcites]{biblatex}
\bibliography{refs}

\usepackage{todonotes}
\usepackage[capitalize]{cleveref}
\usepackage{booktabs}
\usepackage{multirow}

\usepackage[acronym, shortcuts, nohypertypes={acronym}]{glossaries}
\newacronym{AI}{AI}{artificial intelligence}
\newacronym{DL}{DL}{deep learning}
\newacronym{DNN}{DNN}{deep neural network}
\newacronym{RIR}{RIR}{room impulse response}
\newacronym{RL}{RL}{reinforcement learning}

\newcommand{\ie}{i.\,e.,}
\newcommand{\eg}{e.\,g.,}
\DeclareMathOperator*{\argmax}{arg\,max}

\begin{document}

\title{A conceptual framework for learning to listen by reward: Curiosity-driven search for novel sources\\
}

\author{
  \textbf{Andreas Triantafyllopoulos}$^{1,2}$, \textbf{Jakub Šťastný}$^{1}$, \textbf{Alexios Terpinas}$^{1}$,\\\textbf{Tianyi Liu}$^{1}$, \textbf{Yuanqi Wang}$^{1}$, \textbf{Björn W. Schuller}$^{1,2,3,4}$\\
  $^1$CHI -- Chair of Health Informatics, Technical University of Munich, Munich, Germany\\
  $^2$MCML -- Munich Center for Machine Learning, Munich, Germany\\
  $^3$MDSI -- Munich Data Science Institute, Munich, Germany\\
  $^4$GLAM -- Group on Language, Audio, \& Music, Imperial College, London, UK\\
  \texttt{andreas.triantafyllopoulos@tum.de}
}


\maketitle

\begin{abstract}
Reinforcement learning is a powerful learning paradigm that has spearheaded progress in numerous domains.
Its core promise lies in learning through high-level goals without the need for granular labels.
However, it still remains elusive in the realm of audio, where it has received substantially less attention than in computer vision or other domains.
The key question remains: how can agents learn to listen purely via reward-driven exploration?
In this contribution, we present an overview of previous attempts and a new conceptual framework for learning to listen by reward.
Our approach depends on the continuous search for novel sound sources.
We formulate our framework, discuss open technical challenges, and present a first proof-of-concept implementation that showcases the feasibility of our approach.
\end{abstract}

\section{Introduction}
\label{sec:intro}
\Gls{RL} has been pivotal for the widespread adoption of \gls{AI} since the early 2010's, with roots dating back to the early days of \gls{AI} research~\citep{Sutton98-RLA}.
Beginning with the mastery of games~\citep{Mnih13-PAW}, where it notably beat the world's best \emph{Go} players~\citep{Silver16-MTG}, its successes played a pivotal role in bringing \gls{AI} to the spotlight for the general public.
These quickly propagated to neighbouring fields: autonomous driving~\citep{Kiran21-DRL}, robot navigation~\citep{Kober13-RLI}, mathematical reasoning~\citep{Yang24-FMR}, and even the alignment of language models~\citep{Ouyang22-TLM} are all driven largely by \gls{RL}.

Yet, the one domain which has consistently escaped its clutches is audio analysis.
Previous attempts have mainly focused at using \gls{RL} methodologies to optimise \gls{DL} models for metrics that are not differentiable (and thus not amenable to classic optimisation via gradient descent).
Most recently, this was done for speech emotion recognition~\citep{Rajapakshe22-ANP} and audio captioning~\citep{Xu20-ACB, Mei21-AEB, Xu23-BTS}.
While useful for obtaining better performance, this attempt is lacking all the main ingredients of the classic \gls{RL} framing -- that of an \emph{agent} embedded in an \emph{environment} which is \emph{rewarding} it for its \emph{actions}~\citep{Sutton98-RLA}.
This formulation has been partially pursued in the context of navigation and separation; we review relevant works in \cref{sec:related}.
However, the field is still lacking a transparent conceptual framework for \emph{learning to listen by reward}, as well as an overview of the necessary technical considerations for developing \gls{RL} agents for audio.

Reward-based learning for audio holds several advantages over alternative learning paradigms, such as classic supervised or self-supervised approaches.
First of all, it helps to circumvent the relative dearth of data compared to other fields; by utilising known sources and simulation software, it is possible to generate a far greater amount of environments than is currently available in contemporary datasets.
More importantly, \gls{RL} may constitute a pathway towards more generalist audio foundation models.
Contemporary foundation models are often trained a) in self-supervised fashion~\citep{Liu22-ASL}, or, b) as part of audio captioning pipelines~\citep{Triantafyllopoulos25-CAF}.
While these approaches have already led to tremendous progress, goal-driven exploration holds the promise of \emph{continual learning}; by following high-level, prescribed goals (or even self-selected ones), agents can adapt to novel environments and continue learning on-the-fly, a feat which is not easily supported by other paradigms.
Finally, as reward-based learning has interesting connections to how human toddlers learn to listen, pursuing a similar paradigm for machine listening provides a lens to study how the environment, the affordances of an agent, and the rewards they receive shape their listening capabilities.
For all these reasons, \gls{RL} for audio is an important research frontier that has so far remained under-represented.

This contribution aims to bridge this gap in current research by reviewing recent literature (\cref{sec:related}), presenting a coherent formulation for learning to listen by reward (\cref{sec:rl}), and demonstrating the main operating principles by presenting proof-of-concept experiments in \cref{sec:results}.

\section{Related Work}
\label{sec:related}

As mentioned, research on the intersection of audio and \gls{RL} is relatively limited.
Most of the authors use audio as an auxiliary modality to improve the audio-visual navigation capabilities of some robotic agent.
Given the complexity of online learning in real-life scenarios -- {\ie} with robots operating in the physical world -- all of those works rely on simulations.
The most notable of them all, \citet{Chen20-SAN}, introduced the \emph{Soundspaces} dataset -- a large dataset of scanned indoor locations which can be used to simulate sound propagation.
To do so, they extended the \emph{Habitat} simulator with audio synthesis capabilities~\citep{Savva19-HAP}.
The agent's goal is to reach the position of the target source, but it is additionally rewarded in each intermediate step if it moves closer towards the target, {\ie} it is not ignorant of the target's position during training.
This framework has been used by several other works~\citep{Majumder21-MAA, Chen22-S2A, Mo23-AUA} and also extended to moving sources~\citep{Younes23-CMI}.
Similarly, \citet{Hegde21-ATL} utilised the \emph{ViZDoom} environment to add audition capabilities to an agent playing the game of \emph{Doom}; this resulted in agents that were better able to track targets compared to vision-only agents and defeat `deaf' adversaries in a 1-on-1 match-up.

The most similar work to ours is that of \citet{Giannakopoulos21-ADR}, who created a virtual environment using the \emph{Unity} game engine, and used it to simulate audio from two speakers with the goal of reaching one of the two speakers.
This is, to the best of our knowledge, the only work focused exclusively on audio.
However, \emph{Unity} features a very simplistic audio simulation engine with a linear attenuation model and does not account for reflections.
Moreover, their approach hinges on the definition of a `correct' speaker to approach.
This is a more targetted approach than the framework we propose below.
In contrast, our work allows for different simulation software to be used and adopts a more modular objective which does not discriminate between `good' and `bad' targets.

\section{Conceptual Framework}
\label{sec:rl}

\begin{figure}
    \centering
    \includegraphics[width=\columnwidth]{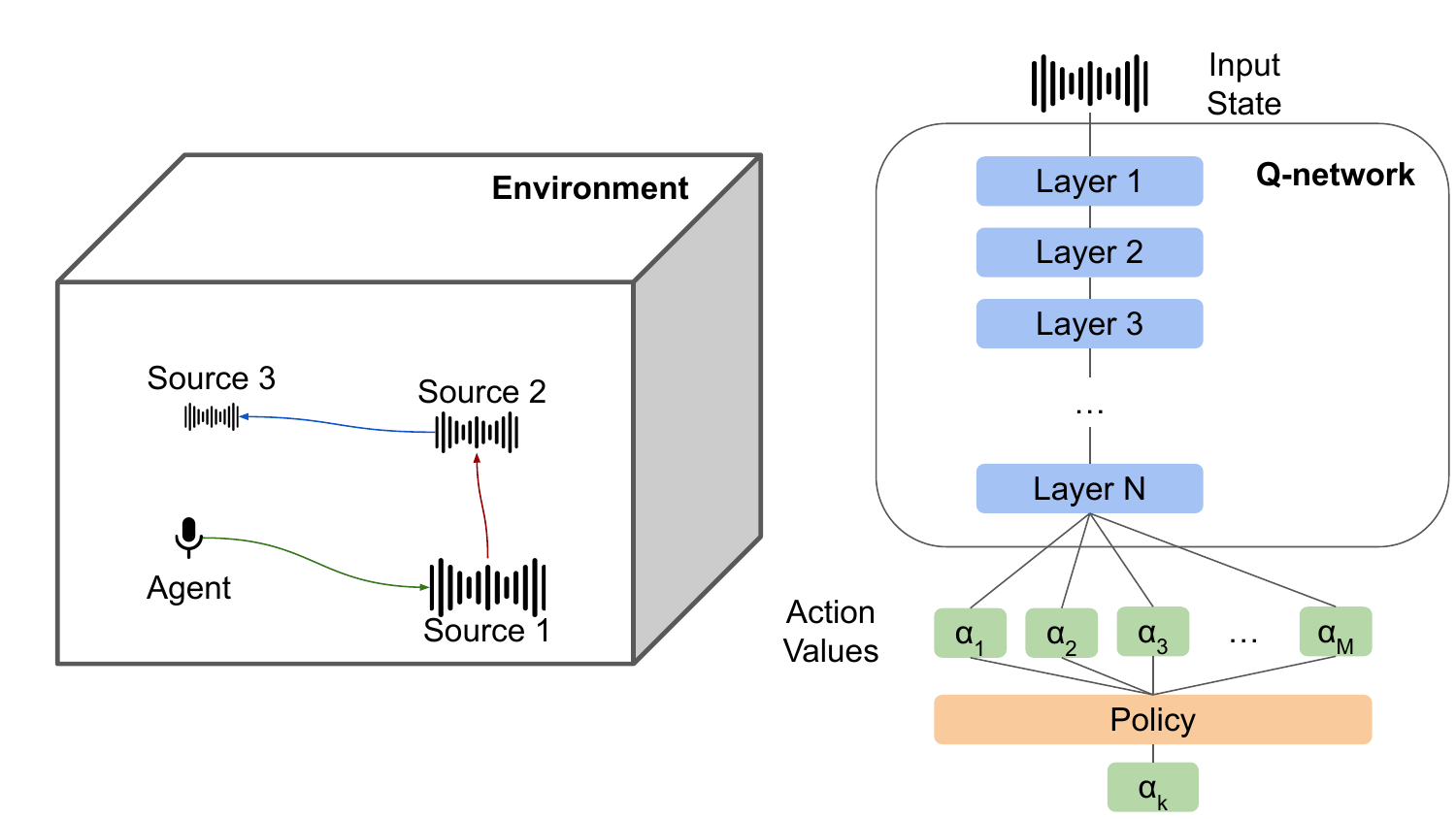}
    \caption{
    Overview of our conceptual framework.
    An agent is searching for novel sound sources in a room by navigating to each of them in sequence.
    At every step, the agent considers the audio it has received from all sources and selects its next action.
    We adopt deep $Q$-learning as our \gls{RL} framework where a $Q$-network is trained to estimate the \emph{value} of each step and a greedy policy for selecting the optimal action.
    }
    \label{fig:framework}
\end{figure}

The notion of an auditory \gls{RL} agent is simple: it constitutes an agent that is able to perceive its environment through listening; it may be also endowed with other affordances, such us vision, but in the present work we limit ourselves to audio as the one and only means to sense the environment.
The main conceptual difficulties thus lie a) in the definition of the environment and b) the designation of a suitable reward function.
We outline our vision for both in the following paragraphs.
We begin by drawing inspiration for how human toddlers learn to use sound for navigation.

\textbf{Motivation from human learning:} Interestingly, there are far fewer studies investigating the importance of sound in the search behaviour of infants~\citep{Bigelow83-DOT, Fazzi11-ROS, Shinskey17-SEM} than the multitudes focused on visual search.
Nevertheless, sound is an important cue that complements the visual modality in the definition of object permanence, such as looking for toys that were hidden from sight but keep generating sound.
Crucially, the importance of sound increases for infants born with severe visual impairments~\citep{Bigelow84-TDO, Fazzi11-ROS}, where they complement touch in the development of abilities such as reaching and crawling.
While few, they still underscore both the importance and the feasibility of using sound for navigation.

\textbf{RL Basics:} \gls{RL} is based on one key principle:
an artificial agent is situated within an environment, takes actions to achieve a particular goal, and is rewarded when that goal is achieved.
The actions that the agent takes are based on the state of the environment as the agent is sensing it.
Crucially, the agent is rewarded for its actions after some indeterminate number of steps; this means that the \emph{value} of each step is not known at the time of execution, but is rather derived at the end of each learning \emph{episode} -- though note that shortcuts are often used, such as the ones discussed for \citet{Chen20-SAN}.

Concretely, we consider agents which interact with the environment in limited horizon episodes.
Each episode comprises a number of discrete \emph{steps}, $\{e\}_0^T$, with $T$ being the maximum number of steps.
At each step, the agent is in a state $s_k$, takes an action, $a_k$, reaches a new state, $s_{k+1}$, and is rewarded for this action by the environment, with the reward in each step denoted as $r_k$.
A learning episode, $E$, is thus defined as a sequence of (state, action, reward) triplets: $E=\{(s_k, a_k, r_k), k \leq T\}$.
A learning algorithm, $A$, accepts as input a set of $N$ training episodes $\{E\}_0^N$ and outputs a \emph{policy}, $\mathcal{\pi}$, for selecting an action at an arbitrary step $k$ given all previous states, actions, and, optionally, rewards: $a_k \sim \mathcal{\pi}(\{(s_j, a_j, r_j), j \in [0, k)\})$.

This policy depends on the estimated \emph{value}, $v_k(s_k)$, of each possible future state.
The key goal of an \gls{RL} algorithm is to determine both the value-function, $Q(\cdot)$, and a policy based on that value function, $\mathcal{\pi}(Q)$.
A typical example of the latter is the $\epsilon$-greedy policy which takes random steps with a probability $\epsilon$ and otherwise selects optimal actions according to the greedy policy~\citep{Sutton98-RLA}.

Our goal is to translate this general \gls{RL} framework to the domain of audio.
The states of the environment, $s_k$, comprise an \emph{auditory stream}.
This auditory stream encapsulates all audio frames, $x_t$, received \emph{up to} that step: $s_k = \{x_t, t \in [0, k)\}$.
The agent processes that stream and uses it to select the next action at time $k$, {\ie} the value function, $f_Q$, depends exclusively on audio input.

\textbf{Embodied agents:} Based on these preliminaries, we can now proceed to define our listening agent.
Inspired by how toddlers use sound for navigation, we consider \emph{embodied} agents which are able to move around in space.
The space of possible actions is then determined by the degrees of freedom that constrain the agent's movement; these can be encoded as simple Cartesian directions (front, back, left, right, up, down) or as spherical coordinates.
Practical considerations require us to \emph{simulate} this environment in order to collect sufficient data for training; however, our ultimate goal is the deployment of these agents in the real, physical world ({\ie} as robots that are able to listen).

\textbf{Audio rewards:} This leaves a last, crucial, open question, namely, how to reward the agent for its actions.
We draw inspiration from how human infants may react to the presence of sounds in their environment -- by navigating towards those sounds and touching their sources -- and reward agents whenever they successfully approach a new sound source in the environment.
Within each episode, we reward the agent for approaching each source only once; this is inspired by how quickly toddlers switch interest to novelty and simultaneously avoids the agent exploiting the first source it encounters to maximise its reward.
We note that this line of approach is similar to sound event detection and localisation~\citep{Politis20-OAE, Mesaros21-SED}.
However, counter to supervised approaches that rely on labelled data featuring sound sources and their relative position to the microphone, reward-based learning instead requires an agent that actively navigates space and is only rewarded after a sequence of steps.

\textbf{Mathematical formulation:} Finally, we formalise our framework in mathematical notation.
We consider an environment $R$ to be a room of fixed size defined in Cartesian coordinates.
Each episode, $E$, begins at time $t=0$ and ends at time $t=T$.
There are $N$ sound sources in the room.
Each sound source $j$ has an initial position $y_j^0$.
The position of each sound source may change at time $t$ if the sources are moving; for non-moving sources, we have $y_j^t = y_j^0, \forall j \in [0, N), \forall t \in [0, T)$.

The agent is defined by its centre of mass, $m^t$, at time $t$.
Its sensing capabilities are limited to a fixed amount of microphones $K$, which are placed in specific locations near its centre of mass.
These microphones capture surrounding audio at a constant \emph{auditory} sampling rate $f_s$.
At each time point, $t$, the agent has access to a $K$-dimensional waveform $\{x(K, \tau), \tau \in [0, t)\}$.
With a fixed \emph{action} sampling rate, $f_a$, the agent determines its next action at time $k$.
The agent receives its reward from the environment after a minimal delay, $\delta$, according to the following rule:
\begin{equation}
    r_k = \begin{cases}
        r_+, \exists j: d(m^{t+\delta}, y_j^{t+\delta}) \leq \epsilon, j \notin \mathcal{F}\\
        r_-, \text{otherwise,}
    \end{cases}
\end{equation}
where $\epsilon$ is a small tolerance value for `touching' a particular sound source, $d$ is a distance function ({\eg} Euclidean), $\mathcal{F}$ is the set of sources that have been previously found, and $r_+$ is the reward given when a new source is found after the last action, whereas $r_-$ is an optional, negative result given after each action where the agent fails to achieve its goal.
The latter is meant to promote exploration and prevent the exploitation of finding a single source and stopping.
The optimal value function is given by the Bellman equation~\citep{Sutton98-RLA}:
\begin{equation}
    \label{eq:bellman}
    Q^\ast(s_k, a_k) = \mathbb{E}(r + \gamma \max_{a'}Q^\ast(s', a'|s_k, a_k)\text{,}
\end{equation}
where $Q^\ast(\cdot)$ is the \emph{optimal} value.
The Bellman equation is interpreted as follows:
the optimal value of each action $a_k$ in each state $s_k$ is given by the reward obtained after performing $a_k$ plus the maximum future expected value (discounted by a weight factor $\gamma<1$).
This maximum expected value, $\max_{a'}Q^\ast(s', a'|s_k, a_k)$, is the expected value if only optimal actions are made following $a_k$.
The Bellman equation is recursive and thus needs to be approximated.


\textbf{Learning algorithm:} In principle, any \gls{RL} algorithm may be used to train the agent~\citep{Sutton98-RLA}.
In the \gls{DL} era of \gls{RL}, the most widely-used algorithm is deep $Q$-learning~\citep{Mnih13-PAW, Fan20-ATA}, a variant of the classic $Q$-learning algorithm which models $Q$-values using a network, $f_Q$, that is trained via gradient descent.
Deep $Q$-learning approximates \cref{eq:bellman} by representing $Q^\ast$ using a \emph{target network}, $f_{\hat{Q}}$~\citep{Mnih13-PAW}.
Specifically, the $Q$-loss used to update the parameters $\pmb{\theta}_Q$ of $f_Q$ given the action $a_k$ taken at state $s_k$ and yielding reward $r_k$ is:
\begin{equation}
    \label{eq:loss}
    L_k = (r_k + \gamma \cdot max_a f_{\hat{Q}} (s_{k+1}, a_k) -  f_Q (s_k, a_k))^2\text{.}
\end{equation}
\begin{figure*}[t]
    \centering
    \includegraphics[width=.33\textwidth]{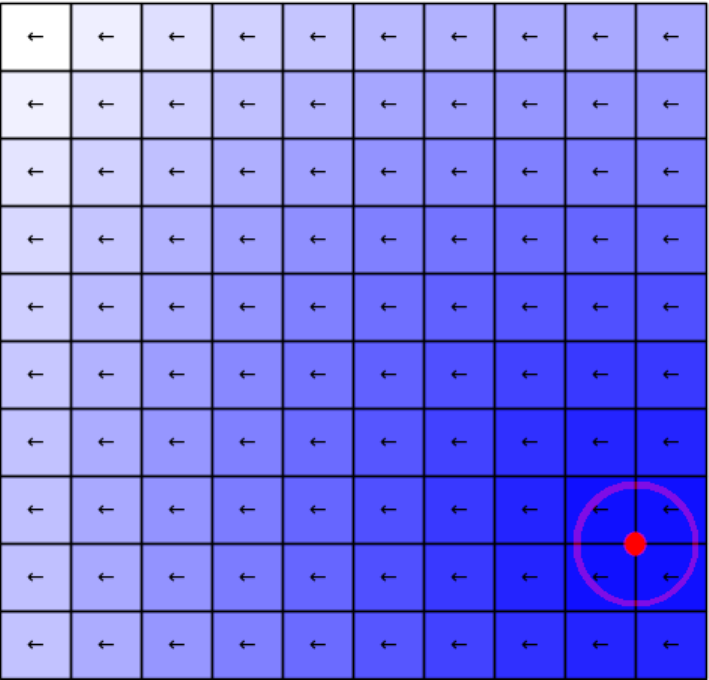}~%
    \includegraphics[width=.33\textwidth]{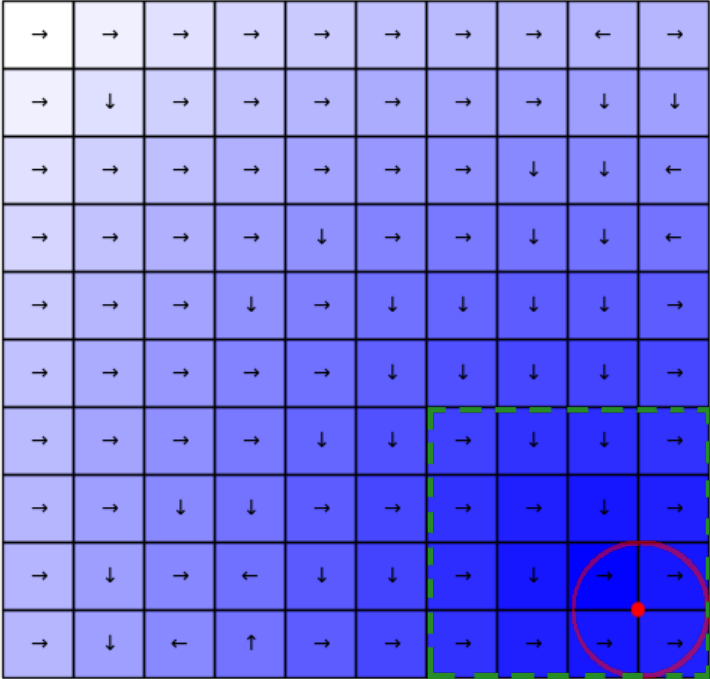}~%
    \includegraphics[width=.33\textwidth]{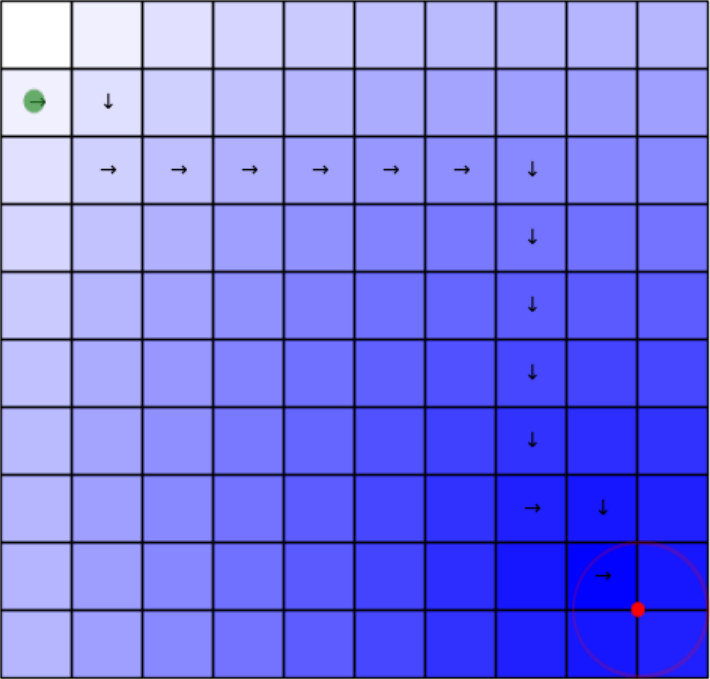}
    \caption{
    Optimal action ($\argmax(f_Q(s_k)$) for random (left) vs trained model (right).
    Arrows designate the direction in which the agent would move if it reached a particular point in the grid.
    Source designated by red dot; red circle is the radius within which the source is considered found.
    Green dashed lines indicate quadrant left out for evaluation.
    Right panel shows one particular trajectory.
    }
    \label{fig:states}
\end{figure*}
The target network is a delayed version of the main $Q$-network, $f_Q$, and is used as a proxy of the optimal $Q$-value, $Q^\ast$.
While $f_Q$ is trained using gradient descent, the parameters of $f_{\hat{Q}}$ are updated at regular intervals to match the parameters of $f_Q$, thus being a `delayed' version of it.
In practice, it may be updated either using `hard' updates, where the parameters $f_{\hat{Q}}$ are set to be an older version of $f_Q$ ({\eg} stored before the last gradient descent update), or using `soft' updates, where the parameters of $f_{\hat{Q}}$ are interpolated based on previous versions of $f_Q$.
Concretely, if the parameters $\pmb{\theta}_{\hat{Q}}$ of $f_{\hat{Q}}$ are updated at step $l$, then, the hard update rule becomes $\pmb{\theta}_{\hat{Q}}^l = \pmb{\theta}_{Q}^{l-r}$, where $r$ is an optional \emph{delay} to avoid mode collapse (if $\pmb{\theta}_{\hat{Q}} = \pmb{\theta}_{Q}$ then, the temporal difference in \cref{eq:loss} becomes $0$).
Accordingly, the soft update rule is implemented as $\pmb{\theta}_{\hat{Q}}^{l} = g(\pmb{\theta}_{\hat{Q}}^{l-1}, \pmb{\theta}_{Q}^l)$, where $g$ is an update function, {\eg} the exponential moving average.

Finally, training data from previous episodes are collected in an \emph{experience replay buffer}~\citep{Lin92-SRA}, which is continually updated with newer episodes.
This buffer contains $(a_k, s_k, r_k)$ triplets with the different states visited by the agent.
It is continuously updated by removing older data and incorporating new ones as the agent learns from experience.
While the simplest update strategy is to follow a first-in-first-out scheme, we found this to result in slower learning due to the sparseness of rewards in our environment.
Our solution was to first remove episodes which did not result in a successful outcome (finding the target source); if the size of the buffer was still exceeded, we then dropped episodes in a first-in-first-out strategy.
    
\textbf{Temporal considerations:} Note that there are two distinct \emph{sampling rates} in our formulation.
The first one, denoted with the index $f_s$, is the granular sampling rate used to characterise audio systems; this is the rate of a classic analogue-to-digital converter that is sampling the audio from a microphone; {\ie} 16\,kHz or higher.
The second one, denoted with the index $f_a$, is the much coarser sampling rate at which the agent is making its decisions; for an embodied agent, this is limited by practical considerations, such as its processing capabilities or the reaction time of its motors.

\textbf{Simulation software:} Given our reliance on simulations, the fidelity, ease-of-use, and execution speed of simulation frameworks becomes an important decision point.
The most widely-used, generic package to simulate \glspl{RIR} is \emph{pyroomacoustics}~\citep{Scheibler18-PAP}.
It supports both image-source model and ray-tracing simulations for both shoebox rooms and arbitrary meshes.
One alternative is \emph{gpuRIR}, which features GPU acceleration but only supports the image-source model~\citep{Diaz21-GAP}.
\emph{Habitat}~\citep{Savva19-HAP} allows for the simulation of arbitrary room geometries with ray tracing; however, we note that the implementation is highly experimental, as it is only available via an abandoned branch of the main \emph{Habitat} repository while its core audio dependency is archived by the project maintainers\footnote{\url{https://github.com/facebookresearch/rlr-audio-propagation}} and contains a binary without accompanying source code.

The major downside of all these simulators is that they are unable to handle moving microphones and/or sources.
This is a seriously limiting factor which impacts the realism of the simulation.
Given that the agent is constantly moving -- and that sources might be moving too -- a realistic simulation would update the \gls{RIR} in each frame, or, at the very least, on a frequent basis.
Given that this would incur a high computational penalty to our algorithm, as we would need to regenerate the \gls{RIR} anew every $\frac{1}{f_s}$ seconds, where $f_s$ is the sampling rate of the target audio, we settled for an approximation that is also followed by prior work~\citep{Chen20-SAN} -- that of synthesising the whole audio from the source at every new position of the agent.
This is tantamount to having a source that repeats itself every $\frac{1}{f_a}$ seconds (the period of the action-taking module), with the agent waiting for the entire audio clip to finish before taking their next action.
While simplistic, this approach is conceptually similar to a toy hidden from a human toddler that repeats a given sound on a frequent basis~\citep{Bigelow83-DOT, Bigelow84-TDO}; given how early we are in the development of \gls{RL} agents that listen, it seems appropriate to begin from a similar basis.

Finally, we note that game engines which allow audio propagation, such as \emph{Unity}~\citep{Giannakopoulos21-ADR} and \emph{ViZDoom}~\citep{Kempka16-VAD} typically feature very simplistic audio propagation models ({\eg} linear roll-off);
nevertheless, they may offer a fast alternative to the previous simulators.
Moreover, modern methods such as audio neural radiance fields aim to approximate \glspl{RIR} using trained neural networks~\citep{Lan24-AVR, Ratnarajah22-MNA}.
Such approaches might prove useful in the future but are not yet mature enough and remain slower and less accurate than the acoustic simulators they are trained on~\citep{Lan24-AVR}.


\section{Experimental Results}
\label{sec:results}

\begin{table}[t]
    \centering
    \caption{Success rate on $1,000$ trials with a max of $50$ steps.}
    \label{tab:res}
    \begin{tabular}{c|ccc}
    \toprule
    $Q$-network & \textbf{Accuracy} & \textbf{Reachability} & \textbf{Reward}\\
    \midrule
    \emph{Random} & 41\% & 08\% & -.89 \\
    \emph{CNN6} & 68\% & 36\% & .08 \\
    \emph{CNN-Transformer} & \textbf{74\%} & \textbf{52\%} & \textbf{.89} \\
    \bottomrule
    \end{tabular}
\end{table}

We created a proof-of-concept implementation for our framework, which relies on \emph{pyroomacoustics}.
We simulated a shoebox room of dimensions $10 \times 10 \times 5$\,m$^3$, with $(0, 0, 0)$ set at the bottom left corner.
The agent was equipped with two omnidirectional microphones, placed at $(\pm.25, \pm.25, 0)$ relative to the centre of mass of the agent.
It was always placed in a constant height ($2.5$\,m) and allowed to move freely in the $(x,y)$ plane with a step size of $.5$\,m.
The source was placed slightly higher ($2.6$\,m) to avoid an exact co-occurrence of the two as this caused errors in the simulation.

The room was divided in $4$ quadrants, with the first three (top left, top right, bottom left) used for training and the last one (bottom right) reserved for evaluation.
During training, both the agent and the source were placed randomly in the $(x,y)$ plane of the three training quadrants.
During testing, the source was always placed in the testing quadrant.
We investigated two alternatives for the $Q$-network:
a) a \emph{memoryless} \emph{CNN6}~\citep{Kong20-PLS} which predicts the next action based on the current state alone.
b) a \emph{CNN-Transformer} model which keeps track of previous states and actions as follows;
the model keeps a memory of the past $7$ states and the corresponding actions;
all $8$ states (past and current) are independently passed through a \emph{CNN6} encoder;
a fixed dictionary is used to create embeddings from each past action and these are concatenated to the state embeddings; 
the final embeddings are passed to a multi-head self-attention layer ($8$ heads) and averaged;
a final linear layer predicts the next action.
\emph{CNN6} was trained \emph{online} for $30$ epochs, whereas $15$ epochs were found sufficient for \emph{CNN-Transformer}.

The $Q$- and $\hat{Q}$-networks were kept constant within each epoch.
The experience replay buffer had a memory of $4,000$ triplets.
At the end of each epoch, the $Q$-network was trained to minimise \cref{eq:loss} using the \emph{Adam} optimiser with a batch size of $64$ and a learning rate of $.0001$.
The experience replay buffer was sampled without replacement $150$ times, using `hard' updates for $\hat{Q}$ every $15$ iterations and setting its value to a delayed version of $Q$ with a fixed delay of $15$ iteration steps.
The reward for reaching the source ($r_+$) was set to $1$.
The agent was assumed to reach the source when their Euclidean distance was $<.6$\,m.
A small negative reward $r_-$ was given for every step where the agent failed to reach the source ($-.1$) and a larger one when it stepped out-of-bounds $-1$.
For exploration, we used the $\epsilon$-greedy strategy, with $\epsilon$ initialised at $.6$ and gradually annealed to $.95$ at the end of each epoch with the following rate: $\epsilon^{k + 1} = 1 - (1 - \epsilon^{k}) \cdot .95$.

To evaluate model performance, we computed the following metrics:
a) \emph{Accuracy}: We generated $1,000$ random source-agent positions and measured whether the model was able to select the optimal action, {\ie} the one that would take it closer to the source;
b) \emph{Reachability}: We generated $1,000$ random source-agent positions and measured whether the able was able to reach its target within $100$ steps without clashing into the room walls;
c) \emph{Average total reward}: We generated $1,000$ random source-agent positions and computed the total accumulated reward, where beyond the training rewards ($r_+=1, r_-=-1$) we also gave a \emph{soft} reward for reducing its distance to the source with each action.
This soft reward was computed as $r_{soft}=0.1\cdot (d_{t}-d_{t-1})$, where $d$ is the Euclidean distance between agent and source.

Results are presented in \cref{tab:res}, where we also include a randomly-initialised $Q$-network for reference.
We observe that the stateful \emph{CNN-Transformer} outperforms the memoryless baseline by a large margin.
It selects the optimal action in $74\%$ of all cases and is able to reach the target in $52\%$ of all evaluation episodes, while accumulating a larger reward ($.89$ vs $.08$).
Nevertheless, \emph{CNN6} also performs better than chance and is able to choose an optimal action with an accuracy $68\%$, though that only leads it to reach the target in $36\%$ of all cases.


\section{Conclusion}
\label{sec:conclusion}
We presented a novel conceptual framework for learning to listen by reward.
Inspired from human learning, the agent is rewarded for navigating to novel sources within an environment.
It is an embodied paradigm for learning which goes beyond previous approaches by focusing exclusively on audio and presenting a simple objective that depends on a very simple prior and does not rely on additional heuristics.
We also presented a practical implementation of this framework based on deep $Q$-learning using a single, stationary source and showed that the agent can generalise to source positions unseen in training.
This work lays the foundation for further developing \gls{RL} for the audio domain, which has thus far remained underresearched by the audio community.
We hope that future work will exploit the benefits of reinforcement learning to build generalist agents that can interact and navigate their environment.
Such agents can shape future applications in the domain of autonomous robotics, enhancing situational awareness and enabling goal tracking that goes beyond visual navigation.
A direct extension of our work could explore the use of multiple, optionally moving, sources and a richer palette of simulated environments.

\section*{Acknowledgments}
This work was partially funded from the DFG's Reinhart Koselleck project No.\ 442218748 (AUDI0NOMOUS).

\newpage
\section{References}
\printbibliography[heading=none]

\end{document}